\documentclass[11pt,letterpaper]{article}
\usepackage{natbib}
\usepackage{array}
\usepackage[dvipdfmx]{graphicx}
\usepackage[labelformat=parens,subrefformat=parens]{subcaption}
\usepackage{amsmath}
\usepackage{amssymb}
\usepackage{url}
\usepackage{amsthm}
\usepackage{comment}
\usepackage{multirow}
\usepackage[a4paper,top=30truemm,bottom=30truemm,left=30truemm,right=30truemm]{geometry}
\usepackage{here}
\usepackage{color}
\usepackage{booktabs}

\usepackage{authblk}

\makeatletter
\newcommand{\figcaption}[1]{\def\@captype{figure}\caption{#1}}
\newcommand{\tabcaption}[1]{\def\@captype{table}\caption{#1}}
\makeatother
\newcommand{\ER}{Erd\H{o}s-R\'{e}nyi }

\theoremstyle{definition}

\newcommand{\hpt}{\hspace{12pt}}

\newtheorem{definition}{Definition}

\title{Referral Hiring and Social Network Structure\thanks{
I acknowledge financial support from the Doctoral Student Fellowship at Kobe University.
I am grateful for the valuable comments received from Tamotsu Nakamura, Teruyoshi Kobayashi and Kohei Yamagata.
}}
\author[]{Yoshitaka Ogisu\thanks{E--mail: 197e109e@stu.kobe-u.ac.jp}}
\affil[]{Graduate School of Economics, Kobe University, Kobe, Japan}
\date{\today}

\begin{document}

\maketitle


\thispagestyle{empty}
\begin{abstract}
\noindent
It is well known that differences in the average number of friends among social groups can cause inequality in the average wage and/or unemployment rate.
However, the impact of social network structure on inequality is not evident.
In this paper, we show that not only the average number of friends but also the heterogeneity of degree distribution can affect inter-group inequality.
A worker group with a scale-free network tends to be disadvantaged in the labor market compared to a group with an \ER network structure.
This feature becomes strengthened as the skewness of the degree distribution increases in scale-free networks.
We show that the government's policy of discouraging referral hiring worsens social welfare and can exacerbate inequality.

\vspace{10pt}

\noindent
{\it JEL Classification}: E24, J31, J64;

\noindent
{\it Keywords}: Referral hiring; social network; network structure; inter-group inequality. 
\end{abstract}
\newpage

\setcounter{page}{1}

\section{Introduction}
    
    Referral hiring is one of the most common practices in the US labor market.
    In 2017, 24\% of job seekers used referrals from friends or relatives as one of their primary job search methods.\footnote{The number is calculated by the author with NLSY 97 (data access 4/3/2020).
    }
    Although the usage rate has decreased dramatically, compared to the data shown in \cite{Holtzer1988search}, many people still use referrals for job seeking.

    Social networks facilitate job-to-worker matching in referral hiring and create inequalities among workers and social groups, which is highlighted by literature \cite{rees1966information} and \cite{GRANOVETTER1977347}.\footnote{For comprehensive surveys, see \cite{topa2011labor} and \cite{ioannides2004job}.}
    In the economy with referral hiring, workers with more friends have lower unemployment probabilities than those with fewer friends, and thus earn higher wages \citep{igarashi2016distributional}. 
    Two main reasons lead to this result. 
    First, workers may be more likely to receive job offers with higher wages directly from their friends \citep{mortensen1994personal, calvo2007networks}. 
    Second, workers with more friends may have more outside options, giving them an advantage in wage bargaining \citep{fontaine2008similar,igarashi2016distributional}.
    In either case, it is argued that the number of friends is one of the keys to shaping a hierarchy among workers in a social network.

    While the number of friends is a valid factor, the network structure within a social network could also play a significant role in determining wage and unemployment rates.
    One remarkable example is via the congestion effect caused by social network structures \citep{calvo2005job}.
    As the number of connections in a social network increases, the chances of unemployed people finding an available job via the social network increases.
    Simultaneously, however, the probability that an unemployed one receives multiple job information also increases.
    If the number of connections increases above a certain level, the latter effect dominates the former one.
    Thus, the unemployment rate increases, and the average wage rate decreases because the number of job matches in the group decreases.
    
    Although a moderate number of friends is desirable for workers in the presence of congestion effects, most workers, typically, cannot have such a number of friends.
    Most social networks observed from data are scale-free networks, in which a distribution of connections (called {\it a degree distribution}) has a scale-free nature \citep{newman2018networks}.
    Since the degree distribution has a heavy tail in a scale-free network, workers with a moderate number of friends cannot be a majority.
    Thus, workers in scale-free networks could be disadvantaged compared to those in simple networks such as \ER or regular networks.
    Then, differences in network structure can be a source of inter-group inequality.
    
    In order to deal with scale-free networks composed of workers, \cite{ioannides2006wages} constructs a search and matching model with an arbitrary degree distribution based on the \cite{pissarides2000equilibrium} model.
    In this case, the network can be regarded as \textit{a configuration model}.
    A configuration model is one of the random networks that has a prespecified degree distribution.
    In a configuration model, first, all nodes are given stubs according to the degree distribution.
    Then, the network is created by selecting two stubs with uniform probability and connecting them to make an edge until there are no more stubs.
    It is well known that configuration models are locally tree-like, that is, with no edges that form a triangle of three nodes, so the clustering coefficient of each node becomes zero when the number of nodes is sufficiently large \citep{catanzaro2005generation,newman2018networks}.\footnote{
    See \cite{newman2018networks} for more detail about a configuration model.
    }
    Although this is one of the limitations of the analysis, the model can still roughly capture the economic impact of scale-free networks.
    In scale-free networks, most workers have only a few friends, even if the average number of friends increases due to workers with a very large number of friends.
    In contrast, in \ER and regular networks, the number of friends of most workers is distributed around or at the mean of degree distributions, respectively.
    This difference can cause inequality among these groups through congestion effects.
    
    To introduce referral hiring, we formulate how a worker's friend finds a vacant job and sends the information to the worker.
    Most previous studies that deal with referral hiring in search and matching models have assumed that workers who are already matched to jobs also randomly obtain information about a vacancy in the labor market and pass it on to their unemployed friends \citep[e.g.][]{calvo2004effects,ioannides2006wages}.
    While this assumption can simplify the analysis, it still cannot illustrate the actual process of referral hiring.
    \cite{tassier2008social} uses job networks to depict a more realistic referral hiring procedure.
    If a worker knows the status of a job position, the job position may be located in or near the worker's office.
    Thus, a job network can be created such that job positions are nodes and the positions in the same office are connected by edges.
    In fact, Job A and Job B can be connected by an edge if a worker who matches Job A can observe the status of Job B.
    According to empirical works by \cite{cingano2012people} and \cite{glitz2017coworker}, job networks, or coworker networks, play a significant role in the employment probability of workers.
    Therefore, when describing the referral process, it is plausible to assume that not only social networks but also job networks are used.

    In this paper, we capture the network structural effects that lead to inter-group inequality.
    It is necessary to incorporate into the model not only macroscopic indicators (e.g., the average number of friends) but also microscopic characteristics, that is how workers are connected or how many hub workers exist within each social network, for instance, to discuss inequality among groups.
    The introduction of microscopic characteristics allows us to examine what kind of network structures is beneficial for worker groups in the labor market.
    In particular, we compare workers in relatively simple networks such as \ER or regular networks with workers in scale-free networks that are frequently observed in reality, and investigate the extent to which the unemployment rates differ.

    Combining \cite{calvo2005job} and \cite{ioannides2006wages}, we construct a \cite{pissarides2000equilibrium} type search and matching model.
    \cite{calvo2005job} points out the congestion effects of referral hiring, while \cite{ioannides2006wages} studies the impact of heterogeneous degree distribution on job finding and inequality.
    There are two types of worker groups having different social networks in the economy.
    Workers in different groups face the same job arrival rate in the market, but the different ones via referral since they have different degree distributions in their social networks.
    Due to differences in social networks, these two groups achieve different unemployment rates, resulting in wage inequality between groups via Nash bargaining.
    
    This paper is related and contributes to two kinds of literature.
    The first one is to investigate inequality caused by referral hiring \citep[e.g.][]{montgomery1991social,calvo2004effects,calvo2007networks,lindenlaub2021network}. 
    In these works, \cite{lindenlaub2021network} is the closest to our interest; however, the approaches are different.
    \cite{lindenlaub2021network} conducts game-theoretic analysis in a setting where worker income is determined by their achievement through efforts and information sharing among workers.
    In our model, we construct a search and matching model, and the income is decided by bargaining in which the crucial factor for the wage determination is the unemployment probability of workers.
    
    The second one is to analyze effects of referral hiring on the market with search and matching model \citep[e.g.][]{fontaine2007simple,galenianos2014hiring,horvath2014occupational,galeotti2014endogenous,zaharieva2013social,zaharieva2018optimal,galenianos2021referral,horvath2018social}.
    In particular, \cite{fontaine2007simple} is the most related work to our research.
    In his model, the social network structure is given by a regular network, in which all nodes in the network have the same number of edges.
    He shows that increasing the average number of friends leads to a low average unemployment rate in the economy with referral hiring in a mathematically tractable model.
    In contrast, we identify network structural effects by capturing degree distributions in addition to the average number of contacts.

    Although several results obtained in this paper are in common with the previous studies, there are additional findings to understand the impact of networks on inter-group inequality in the labor market.
    Firstly, the structure of a social network, or how it is connected, can be one of the factors causing inequality among groups.
    Even if two groups have the same average number of contacts and all workers have the same productivity, these groups can achieve different average unemployment rates and wages.
    
    Secondly, worker groups with scale-free networks tend to be worse in unemployment and/or wage rate. 
    In the analysis, we use \ER, regular, and scale-free networks with the same average number of contacts for social networks, to compare the average wage and unemployment rate at the equilibrium.
    Between a group of workers having an \ER network and a regular network, the gaps in unemployment rates are not substantial.
    While, if a worker group holds a scale-free network structure, they suffer from a non-negligible disadvantage in most cases.
    The result implies that workers in scale-free networks are more congested in sharing vacancy information than workers in \ER networks if the average number of contacts is the same.
    Since the empirical observations show that numerous social networks form scale-free network structures, theories based on \ER or regular networks could not be applied directly to empirical works.

    Thirdly, if connectivity in the job network increases, the inequality expands among groups having the same average number of contacts but different social network structures.
    Larger connectivity in the job network leads to a higher probability that a referrer gets vacancy positions, and as a result, workers get a lower unemployment probability.
    However, that benefit is allocated unequally between groups.
    Groups that can share vacancy information more efficiently reap greater benefits.

    Lastly, restrictions on referral can deteriorate not only social welfare, but also inequality.
    We check the effects of referral restrictions on worker groups with different network structures.
    Referral restrictions influence social welfare and inequality between groups.
    On the one hand, social welfare decreases since unemployment rates increase and wage rates decrease in all worker groups.
    On the other hand, the effect on inequality could be changed depending on referral prevalence.
    If the impact on a disadvantageous group is much more than on an advantageous group, the inequality expands.
    In this situation, the result in this paper is consistent with \cite{igarashi2016distributional}.

    This paper is constructed as follows.
    We build the model in section \ref{sec:model}, and solve it numerically in section \ref{sec:numerical_sol}.
    Then we go on comparative analyses including changing job network connectivity and restriction of referral in section \ref{sec:compara} and summarize the discussion in section \ref{sec:conclusion}.

\section{The Model} \label{sec:model}

    We mainly follow the works introducing the social network structure to search and matching model \citep[e.g.][]{calvo2005job, ioannides2006wages,fontaine2007simple}.
    Time is continuous, and the horizon is infinite.
    There is a free entry of jobs, and one job matches one worker.
    We only consider the steady state of the economy, with a fixed number of workers and worker groups.
    Further, there is no on-the-job search in this economy.
    
    Unemployed workers seek a job at random in the labor market; moreover, they can ask for vacancy information from connected contacts in the social network.
    When workers are employed, they sometimes look for vacancy information to watch the state of the adjacent jobs through the job network.
    When the event occurs, they can send its information to all unemployed contacts.

    \subsection{Workers}
        Workers are risk-neutral and maximize expected discounted utility with a discount rate $r>0$.
        All workers have identical productivity.
        Let $L>0$ be the number of workers and sufficiently large.
        Moreover, workers make social groups, indexed by $i  \in \{1,2,\dots ,I \}$, and each worker can belong to only one group.
        Denote the number of workers in group $i$ as $L_i$, then $L = \sum_{i=1}^{I} L_i$.
        
        Job matches break up by job destruction which occurs at constant probability $\delta \in [0,1]$.
        If a worker is employed, the worker serves the labor force to the matched job and receives wage $w_i$.
        While, an unemployed one produces home production $b > 0$, and at the same time, seeks a job.
        A job is an arrival to an unemployed worker with probability $p_i$.
        Let $W_i$ and $U_i$ be the value functions of the employed and unemployed workers in group $i$, they satisfy the following equations;
        \begin{align}
            r W_i &= w_i - \delta ( W_i - U_i ), \label{eq:W_value}\\
            r U_i &= b + p_i ( W_i - U_i ). \label{eq:U_value}
        \end{align}
        
        When workers seek a job, there are two available ways; market search and referral.
        Therefore, arrival rate $p_i$ is also constructed by two parts
        as arrival rate through the market $p_i^M$ and referral $p_i^R$.
        
        We employ the matching function proposed by \cite{pissarides2000equilibrium} for market matching.
        In the market, thus, vacancy information arrives at random.
        Let $u$ and $v$ be unemployment rate and vacancy rate, respectively, and define market matching function as $M(Lu, L v) = \gamma (Lu)^{\eta} (L v)^{1-\eta}\ (\eta > 0)$.
        The matching number of group $i$ is determined proportionally to the number of unemployed workers in group $i$.
        With denoting $u_i$ as unemployment rate of group $i$, the number of unemployment of group $i$ is $u_i L_i$.
        Since the whole number of market matching is $M(Lu, L v)$, the number of matching in group $i$ is $M(Lu, Lv) L_i u_i / L u $.
        Consequently, the job arrival rate through the market is given by, $p_i^M = L_iu_i/Lu \times M(Lu,Lv)/u_i L_i = \gamma \left(u/v \right)^{\eta-1}$.
        
        Unemployed workers can use referrals for job-seeking besides the market search.
        An unemployed worker asks all the contacts to refer any jobs and if at least one contact holds vacancy information, the unemployed person succeeds in job matching through referral.
        For a mathematical description of the referral procedure, we need to denote the degree distributions of the social networks and the job network.
        Assume that there is no edge between workers of different groups.
        Let $d_i \sim F_i(n),\ n \in \mathbb{N}$ be a degree, which is a stochastic variable of the number of contacts for workers in group $i$, where $F_i(d_i)$ is distribution function of group $i$.
        We denote the probability mass function of group $i$ as $f_i (d_i)$.
        Contrary to social networks, we assume that the job network is constructed of a random regular network for simplicity.\footnote{
        If job entry is determined endogenously and the job network makes up more complex networks, the value of the vacant incumbent job would be ambiguous.
        Since the main focus of the paper is the social network structure, we avert the difficulty caused by the heterogeneity from the firm network.
        }
        A random regular network is generated, such that all edges are created at uniformly random among nodes under the constraint that each node has $k \in \mathbb{N}$ degree.
        Then, the network is often called $k$-regular network.
        We simply refer to random $k$-regular networks as regular networks.
        Thus, the degree of the job network can be denoted by $d_f$, which can treat as a deterministic variable.
        
        We denote $P_i$ as the probability that an arbitrary worker in group $i$ holds vacancy information and it makes up as follows.
        Pick up an arbitrary worker in group $i$.
        The worker is employed with probability $(1-u_i)$.
        Suppose that the worker is employed and searches for vacant jobs through the job network.
        Each adjacent job is filled with probability $1 - v / (1-u+v)$, hence, the probability that all connected jobs are already filled is $[1-v/(1-u+v)]^{d_{f}}$.
        Subtracting it from $1$, we get the expected value of an arbitrary worker in group $i$ is employed and connected at least one vacancy job, $(1-u_i) [1-\{1-v/(1-u+v)\}^{d_f}]$.
        However, the referral event occurs only when the employee watches job vacancy information with probability $\phi \in [0,1]$.
        Consequently, we get
        \begin{equation}
            P_i = \phi (1-u_i) \left[ 1 - \left(1 - \frac{v}{1- u +v} \right)^{d_f} \right]. \label{eq:P_i}
        \end{equation}
        In this formulation, $\phi$ can be understood as search frequency used in \cite{fontaine2008similar}, which is the parameter indicating eagerness of workers to seek jobs for their friends.

        An arbitrary unemployed worker with degree $n$ in group $i$ can get at least one vacancy information with probability $1-(1-P_i)^n$.
        Consequently, the job arrival rate of group $i$ through referral is
        \begin{align}
            p_i^R &= \sum_{n} \left[ 1 - ( 1 - P_i)^{n} \right] f_i (n) \nonumber \\
                &= E_i \left[ 1 - ( 1 - P_i)^{d_i} \right], \label{eq:pR}
        \end{align}
        where, $E_i[\cdot]$ is the expectation operator over the degree distribution of the social network made up by workers in group $i$.
        
        As a result, we can write the average firm arrival rate for workers belonging to group $i$ as
        \begin{equation}
            p_i = \gamma\left( \frac{u}{v} \right)^{\eta-1} + E_i \left[ 1 - ( 1 - P_i)^{d_i} \right]. \label{eq:arri_p}
        \end{equation}
    
    \subsection{Jobs}
        For each job, discount profit is maximized.
        If a job is vacant, it searches for a worker in the labor market with a cost of $c$ and via referral without cost.
        Assume that vacant jobs cannot exit from the market deliberately.\footnote{This assumption is needed for the existence of the stable job network.}
        We denote the arrival rate of worker $i$ as $q_i$.
        If a job is filled, it produces an output of $y > b$ and pays a worker wage $w_i$.
        A filled job is destructed with probability $\delta$.
        Let $J_i$ and $V$ be the value functions of filled jobs matched with workers in group $i$ and vacant jobs, then these given by
        \begin{align}
            r J_i &= y - w_i - \delta ( J_i - V ), \label{eq:J_value} \\
            r V &= - c + \sum_i q_i ( J_i - V ). \label{eq:V_value}
        \end{align}
        We must note that $V$ is regarded as the expected value over all worker groups.
        
        Since the number of matched workers in group $i$ is given by $L_i u_i p_i$, which is equivalent to the number of matched jobs, the arrival rate of workers in group $i$ is determined by
        $q_i = L_i u_i p_i / L v$.
        
    \subsection{Equilibrium}
    
        We specify the equilibrium of the economy in this section.
        Firstly, the vacant job value must be zero at the equilibrium;
        \begin{equation}
            V = 0. \label{eq:free_entry_condition}
        \end{equation}
        If the vacant value is positive ($V> 0$), jobs outside continue to enter the market for positive profit.
        In this case, since there are implicitly infinite jobs in this economy, vacant jobs keep increasing, and lastly, the vacant value reaches zero because of competition.
        In contrast, if $V < 0$, there is no firm to enter the market, whereas existing jobs are continuously destructed.
        As a result, $V$ keeps increasing until $V=0$.
        It is often called the ``free entry condition''.
        
        Next, we suppose that the wage for each group $i$ is decided by Nash bargaining in this economy, hence,
        \begin{equation}
            w_i = \max_{w_i} (W_i - U_i)^{\beta} (J_i - V)^{1-\beta}. \label{eq:nash_bargain}
        \end{equation}
        
        Lastly, at the steady state, the inflow into and the outflow from the unemployment pool are equivalent, thus,
        \begin{equation}
            u_i p_i = \delta (1-u_i) \hpt \Longleftrightarrow \hpt p_i = \frac{\delta (1-u_i)}{u_i}. \label{eq:in_out_condition}
        \end{equation}
        
        Here, we define the equilibrium as follows.
        \begin{definition}
            \it
            An equilibrium is a set of $(u_i,u,v,W_i,U_i,J_i,V,w_i)$ such that:
            \begin{enumerate}
                \item The steady state of labor market is given by (\ref{eq:in_out_condition}) for all $i$.
                \item The worker wage is determined by (\ref{eq:nash_bargain}).
                \item Free entry condition is satisfied: (\ref{eq:free_entry_condition}).
            \end{enumerate}
        \end{definition}
        
        For solving the model, let $S_i \equiv J_i - V + W_i - U_i$.
        From (\ref{eq:nash_bargain}), we get $J_i - V = (1-\beta ) S_i$ and $W_i - U_i = \beta S_i$.
        Thus, with (\ref{eq:free_entry_condition}), $w_i$ is determined by
        \begin{equation}
            w_i = y - (r + \delta)(1-\beta) S_i. \label{eq:wage}
        \end{equation}
        
        We can rearrange the conditions of value functions and get
        \begin{equation}
            S_i = \frac{y-b}{r + \delta + \beta p_i}. \label{eq:S_i_from_wage_determination}
        \end{equation}
        
        By (\ref{eq:V_value}), (\ref{eq:free_entry_condition}) and wage bargaining solution, we can derive
        \begin{equation}
            \sum_i q_i S_i = \frac{c}{(1-\beta)}. \label{eq:S_i_from_free_entry}
        \end{equation}

        From (\ref{eq:S_i_from_wage_determination}), (\ref{eq:S_i_from_free_entry}), we obtain
        \begin{equation}
            v = \frac{(y-b) (1-\beta) \delta}{c L} \sum_i \frac{ u_i (1-u_i) L_i}{u_i (r + \delta) + \beta \delta (1-u_i)}. \label{eq:vacancy_rate_ss}
        \end{equation}
        
        At the equilibrium, equation (\ref{eq:in_out_condition}) and (\ref{eq:vacancy_rate_ss}) must be satisfied simultaneously.
        Once the combination of $(u_i, v)$ satisfying them is found, $p_i$ is determined by (\ref{eq:in_out_condition}) and later $S_i$ and $w_i$ are given by (\ref{eq:S_i_from_wage_determination}) and (\ref{eq:wage}), respectively.
        Finally, we attain workers and jobs values from (\ref{eq:W_value}), (\ref{eq:U_value}), (\ref{eq:J_value}) and (\ref{eq:V_value}), therefore, the model is closed.
        We can find the equilibrium with some appropriate parameters as show below.
        
\section{Numerical Solutions} \label{sec:numerical_sol}

\begin{figure}[t]
        \centering
        \begin{minipage}{0.3\hsize}
        \includegraphics[width = \hsize]{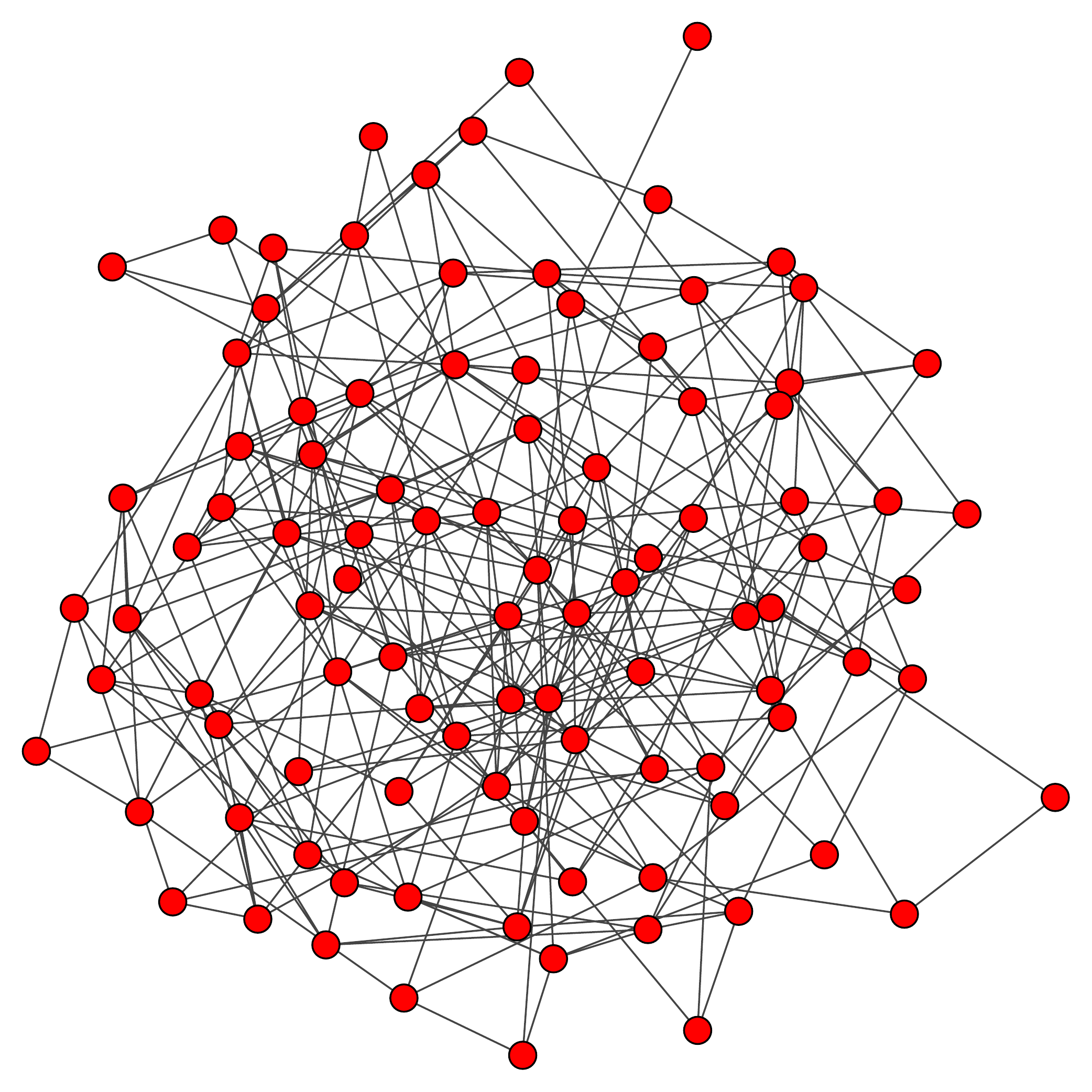}
        \subcaption{\ER network}
        \label{subfig:ER_network}
        \end{minipage}
        \begin{minipage}{0.3\hsize}
        \includegraphics[width = \hsize]{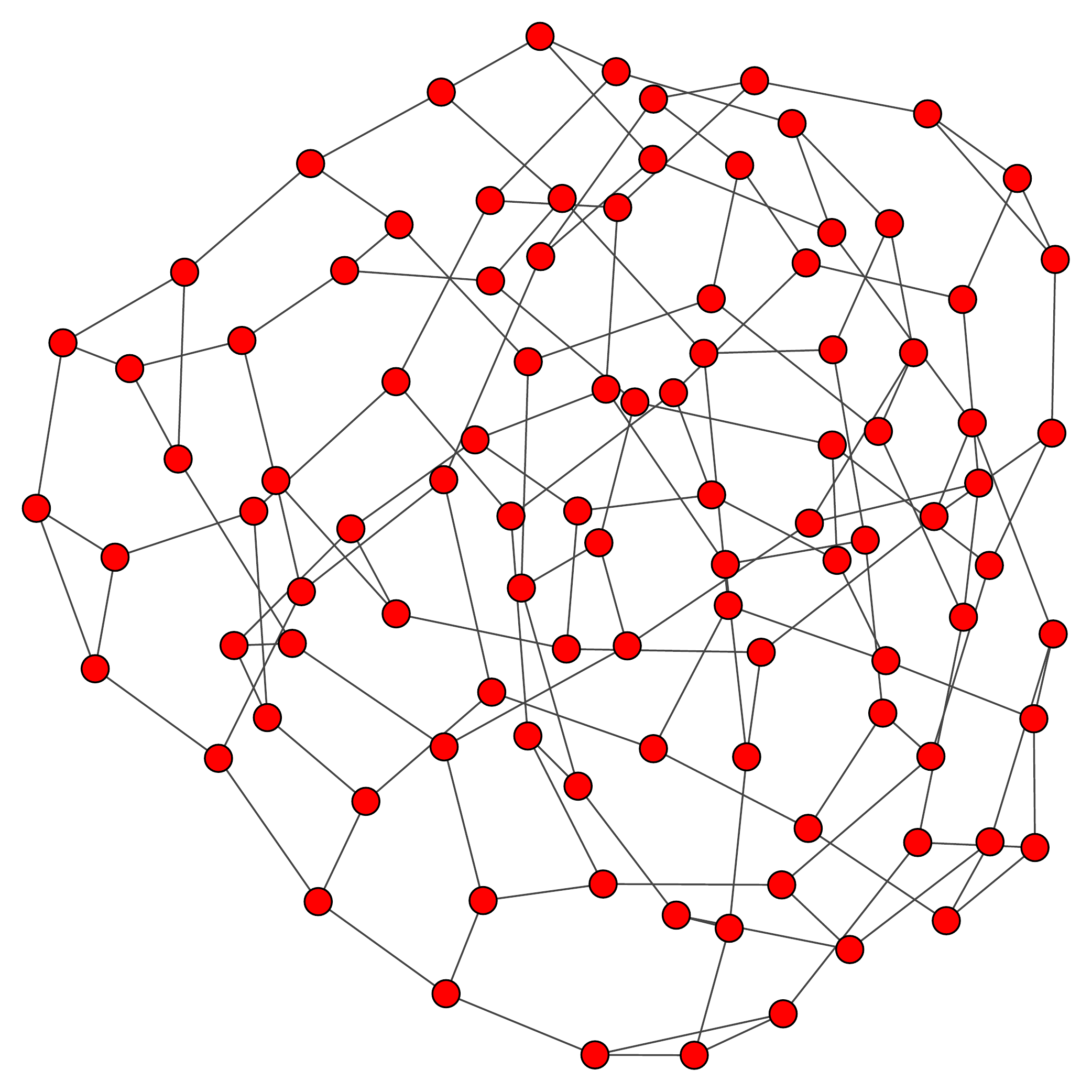}
        \subcaption{Random regular network}
        \end{minipage}
        \begin{minipage}{0.3\hsize}
        \includegraphics[width = \hsize]{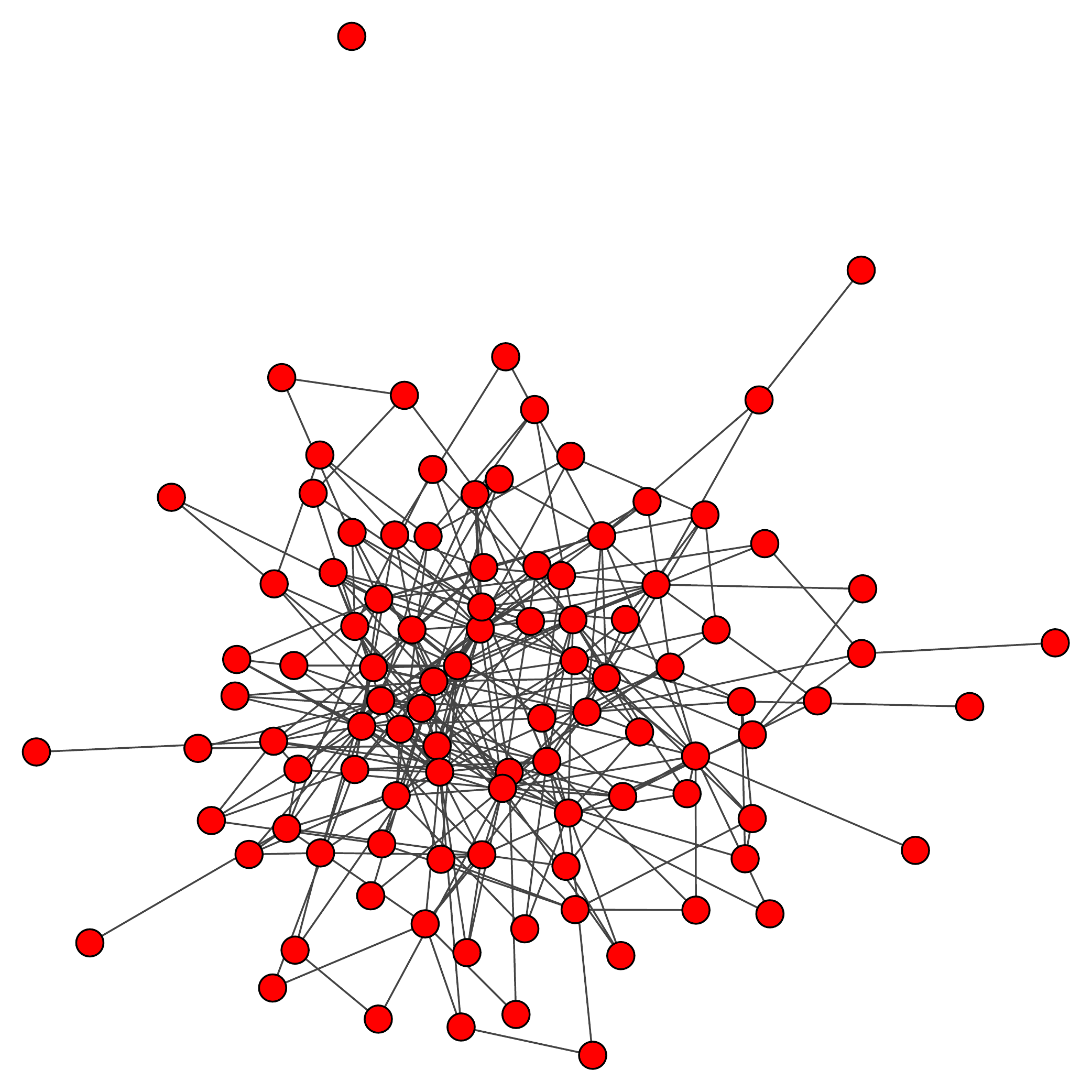}
        \subcaption{Scale-Free network}
        \end{minipage}
        \caption{Network structures with 3 neighbours}
        \label{fig:network_structures}
    \end{figure}

    In this section, we focus on two points to identify the influences of network structures; the average number of contacts in social networks and the difference in network structure.
    The main interest of the paper is the effect of network structures on inequality, while there are many network structures in empirical findings.
    We pick up three representative network structures; \ER, (random) regular, and scale-free networks.
    Examples of these networks are shown in Figure \ref{fig:network_structures}, in which the average number of contacts sets 3.
    An \ER network is a random network with $N>0$ nodes and probability $\pi$ in which any two nodes are connected with $\pi$.
    Then, the all possible combinations are given by $\binom{N}{2}$ and each pair is connected with $\pi$; thus, the degree distribution is given by a binomial distribution, $B\left(\binom{N}{2}, \pi\right)$.
    If $N$ is sufficiently large, it can be approximated by a Poisson distribution.
    In turn, a scale-free network is a network in which the probability of generating edges is not uniform, and the degree distribution has a power law.
    There are various ways to generate scale-free networks.
    In the paper, we only need the information of degree distributions; therefore, we use Zipf distributions for degree distributions of scale-free networks.\footnote{
    Zipf distributions are also called Zeta distributions.
    The probability mass function of a Zipf distribution is $z(k,\alpha) = 1/\zeta (\alpha) k^\alpha$ for $k \ge 1$ and $z(0, \alpha) = 0$, where $k$ is the degree, $\alpha$ is the scale parameter, $\zeta (\cdot)$ is the Riemann zeta function.
    }
    As the scale parameter $\alpha$ approaches $2.0$, the tail of the distribution becomes fatter.
    
    \begin{table}[t]
    \centering
    {\small
        \caption{Parameters}
        \begin{tabular}{clrl} \toprule
             \multicolumn{2}{l}{Given parameters} & Values & Sources \\ \midrule
             $y$ & output & $1$ & normalization \\
             $b$ & home production & $0.4$ & \cite{shimer2005cyclical} \\
             $r$ & discount rate & $0.012$ & \cite{shimer2005cyclical} \\
             $\eta$ & exponent of matching function  & $0.72$ & \cite{shimer2005cyclical} \\
             $d_f$ & number of contacts of a job & $16$ & SUSB \\
             $\delta$ & job destruction rate & $0.036$ & JOLTS \\ \midrule
             \multicolumn{2}{l}{Calibrated parameters} &  & \\ \midrule
             $\gamma$ & market matching efficiency & $0.402$ & \\
             $\beta$ & worker's bargaining power & $0.028$ & \\
             $c$ & vacancy cost & $7.188$ &  \\
             $\phi$ & referral frequency & $0.048$ &  \\ \bottomrule
            \end{tabular}
    \label{tab:calibration_parameter_fix}}
    \end{table}
    
    For numerical calculations, we inherit the parameters in \cite{shimer2005cyclical}, and set $y=1$ for normalization, and $b=0.4$, $r=0.012$ and $\eta=0.72$.
    For the job destruction rate, we put $\delta=0.036$ based on the 2017 average of the separation rate in the Job Openings and Labor Turnover Survey (JOLTS).
    Further, we set $d_f= 16$ because the number of employees per establishment is about $16$ based on the calculation from 2017 SUSB Annual Data Tables.\footnote{\url{https://www.census.gov/data/tables/2017/econ/susb/2017-susb-annual.html} (accessed on July 2022)}
    We use this number as an approximation for the number of coworkers, also the number of observable job positions for a single worker.
    $E_i[d_i]$ will be set for each group later.
    There are 2 groups in the economy and set group sizes equivalent ($L_i=10^6\ \forall i$).\footnote{
    The sizes of worker groups do not affect unemployment rates and wages in the setting.
    Each group size is a critical factor to determine other indicators (e.g., Gini coefficient).
    However, all groups are analyzed with equal network size, and the size effect is not discussed in this paper.}
    
    The rest parameters, $\gamma$, $\beta$, $c$, and $\phi$, are calibrated.
    In the baseline case, assume that all worker groups have \ER network structures and the same expected degree.
    We set the average degree $E_i[d_i] = 22.47~\forall i$ based on the data provided by International Social Survey Programme (ISSP): Social Networks and Social Resources, 2017.
    The calibration targets at the equilibrium are as follows:
    \begin{quote}
        \begin{enumerate}
            \item The average unemployment rate is $4.4\%$ (U.S. Bureau of Labor Statistics).
            \item The average value of the unemployed per job opening is $1.1$ (JOLTS). 
            \item The share of labor compensation in GDP is $0.6$ (Penn World Table 10.0).
            \item $50\%$ of workers finds jobs by referral \citep{igarashi2016distributional}.
        \end{enumerate}
    \end{quote}
    The first and second targets mean $u = 0.044$ and $u/v=1.1$, resulting in $v=0.04$.
    The third target indicates $w_i=0.6$ in the model.
    The last target implies $p_i^R/p_i = 0.5$.
    These targets lead to $(\gamma, \beta, c, \phi) = (0.402,0.028,7.188,0.048)$.
    We summarize parameters in Table \ref{tab:calibration_parameter_fix}.
    For comparing results, we also calculate the Gini coefficient and social welfare defined by $SW \equiv \sum_i [ y (1- u_i) + b u_i ] L_i/L - c v$.

    \subsection{Effects of Average Contacts}
    
    At first, we check the effects of the average degree in each network structure case.
    Set one group $E_1[d_1]=15$ and the other group $E_2[d_2]=30$, and calculate the unemployment rate, wage, and Gini coefficient at the equilibrium.
    Results are shown in Table \ref{tab:effects_of_average_degree}.\footnote{
    For a scale-free network case, we choose each scale parameter $\alpha_1, \alpha_2$ such that the average degree of the distribution adjusted by each target value, $(E_1[d_1],E_2[d_2]) = (15,30)$.}

\begin{table}[t]
    \centering
    \caption{Effects of the average contact for each network structure}
        {\small
        \begin{tabular}{ccccc} \hline
             Network Structure & \ER & Regular & Scale-Free \\ \hline
             $(u_1, u_2)$ (\%) & $(5.10, 3.94)$ & $(5.08, 3.93)$ & $(8.14, 8.10)$  \\
             $(w_1,w_2)$ & $(0.581, 0.615)$ & $(0.582 , 0.615)$ & $(0.529, 0.529)$  \\
             Gini & $ 1.451\times 10^{-2}$ & $1.454 \times 10^{-2}$ & $2.310 \times 10^{-4}$ \\
             $SW$ & $0.685$ & $0.685$ & $0.627$ \\
             \hline
        \end{tabular}
    \label{tab:effects_of_average_degree}}
\end{table}

        In each case, the inequality between groups is generated by the different expected degrees, consistent with \cite{fontaine2007simple}.
        If the social network structures are the same, the worker group with a greater average degree can achieve a lower unemployment rate on average.
        Furthermore, the lower unemployment rate can give the group a higher average wage because the worker has better outside options.
        
        Next, we move on to a comparison among cases.
        For the Gini coefficient, it is highest in the \ER network case and lowest in the scale-free network case.
        However, in the scale-free case, unemployment rates are higher than in the \ER and regular, leading to the social welfare being worst.
        The differences are due to network structural features.
        We see what structure is better for worker groups in the next section.
        Then, we focus only on the unemployment rate since a lower unemployment rate leads to a higher wage.

    \subsection{Effects of Network Structure}

        The next step analyzed in the section is how much the network structural difference affects inequality.
        One way to check the effects is to compare the two groups with the same expected degree while differing in network structures.
        First of all, let the social networks of group 1 and group 2 be an \ER network and a regular network, respectively.
        Then, for various expected degrees, we calculate the unemployment rate for each group at the equilibrium.
        
\begin{figure}[t]
        \begin{minipage}{0.46\hsize}
            \centering
            \small
            \includegraphics[width = \hsize]{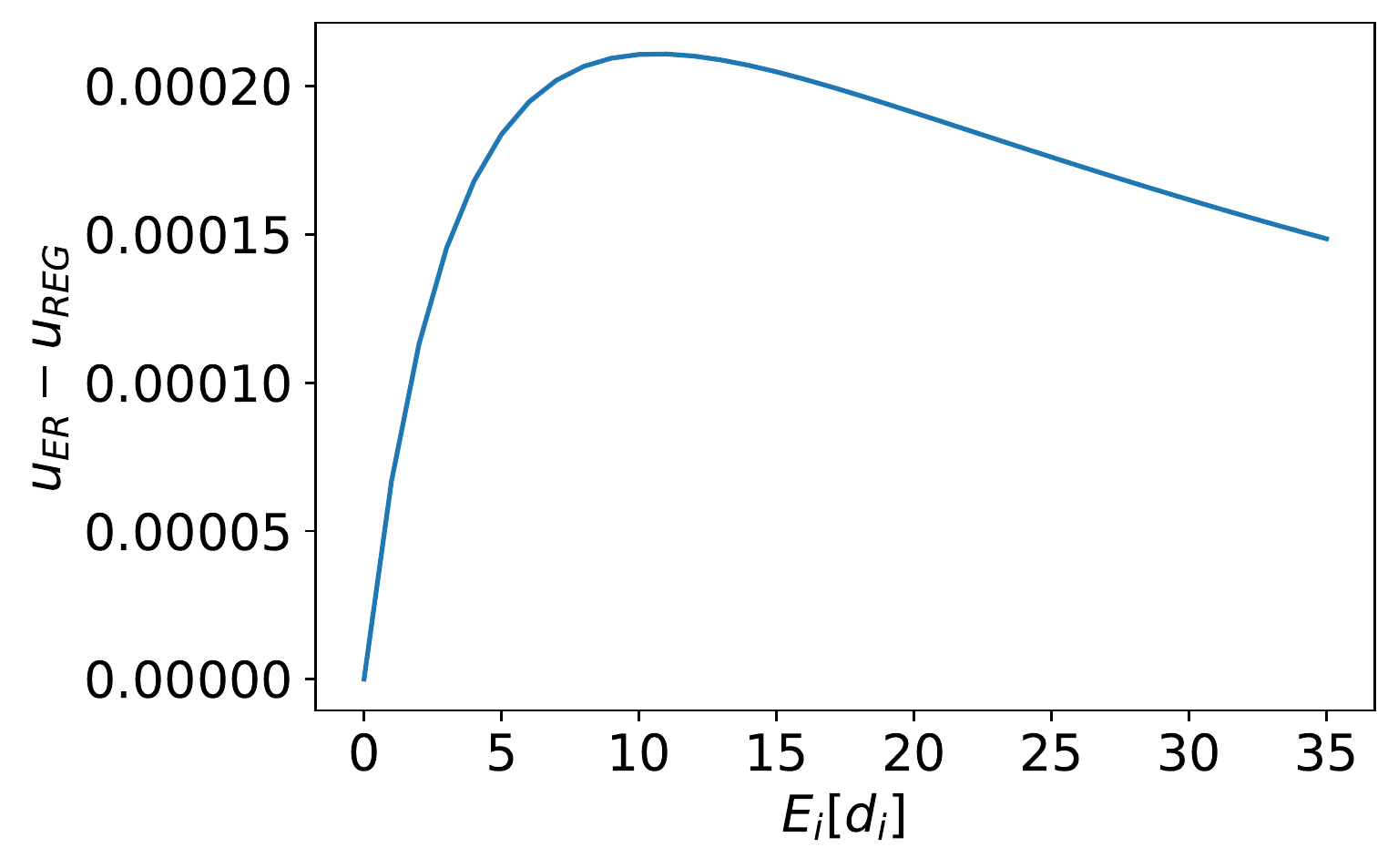}
            \subcaption{$u_{ER}-u_{REG}$}
            \label{fig:gap_random_vs_regular}
        \end{minipage}
        \begin{minipage}{0.5\hsize}
            \centering
            \small
            \includegraphics[width = \hsize]{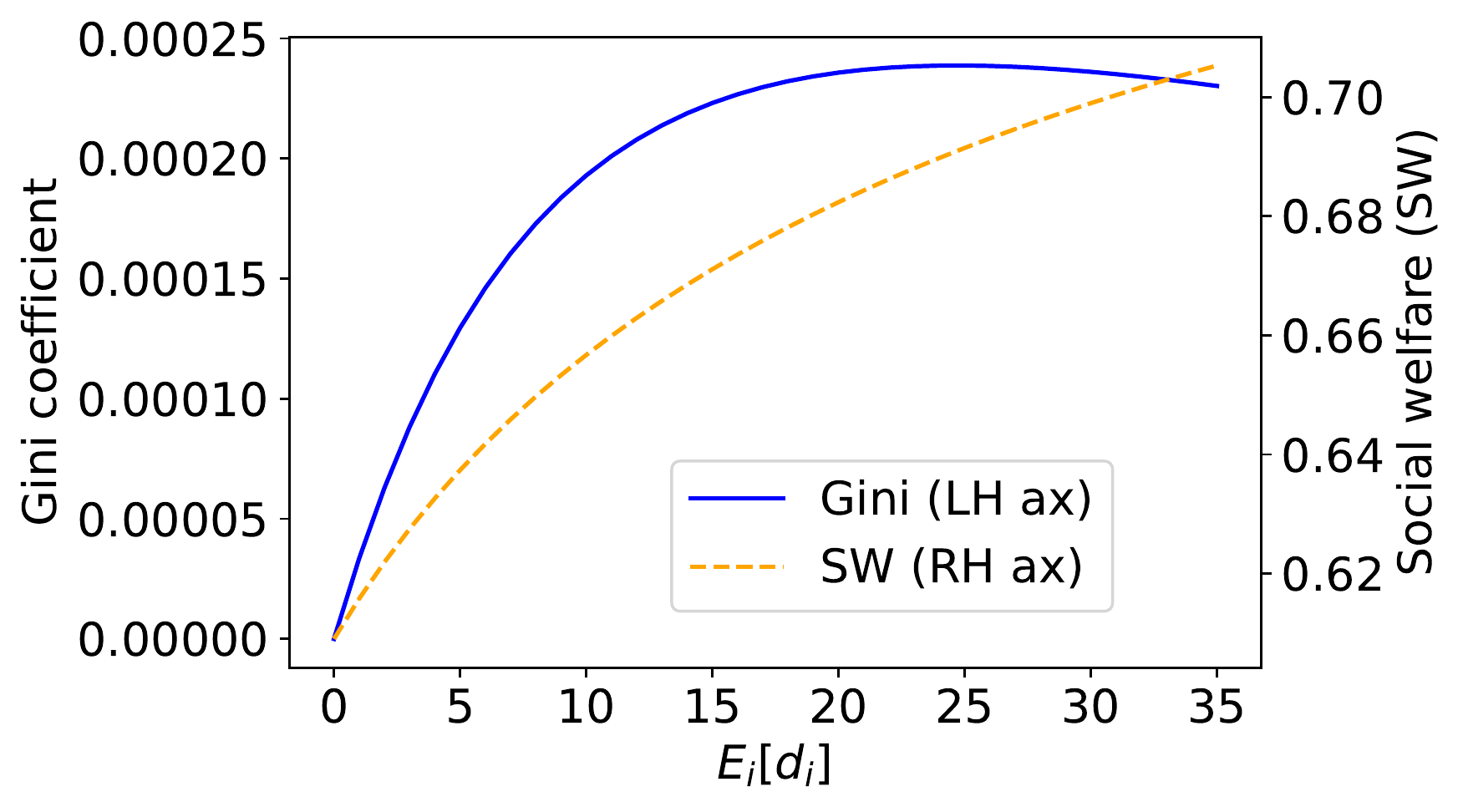}
            \subcaption{Gini coefficient and social welfare}
            \label{fig:random_vs_regular_G}
        \end{minipage}
        \caption{Comparison of unemployment rate, Gini coefficient, and social welfare. \ER network group ($ER$) v.s. regular network group ($REG$).}
        \label{fig:random_vs_regular}
    \end{figure}

        We show the gap of unemployment rate between group 1 (\ER network) and group 2 (regular network) in Figure \ref{fig:gap_random_vs_regular}, and Gini coefficient and social welfare in Figure \ref{fig:random_vs_regular_G}.
        Obviously, there is no inequality at the $E_i [d_i] = 0$ and in $E_i [d_i] > 0$, the inequality begins to rise and it reaches maximum value around $E_i [d_i] = 25$.
        On the contrary, social welfare monotonically increases in $E_i[d_i]$.
        
        Under the setting, in a regular network, workers have the same number of contacts.
        By contrast, in an \ER network, some workers have a lower degree than $E_i [d_i]$, and their unemployment probabilities are usually higher than those who have the degree $E_i [d_i]$.
        The inequality between the two groups could be caused by the number of workers who have a high unemployment probability.

    \begin{figure}[t]
        \begin{minipage}{0.43\hsize}
            \centering
            \small
            \includegraphics[width = \hsize]{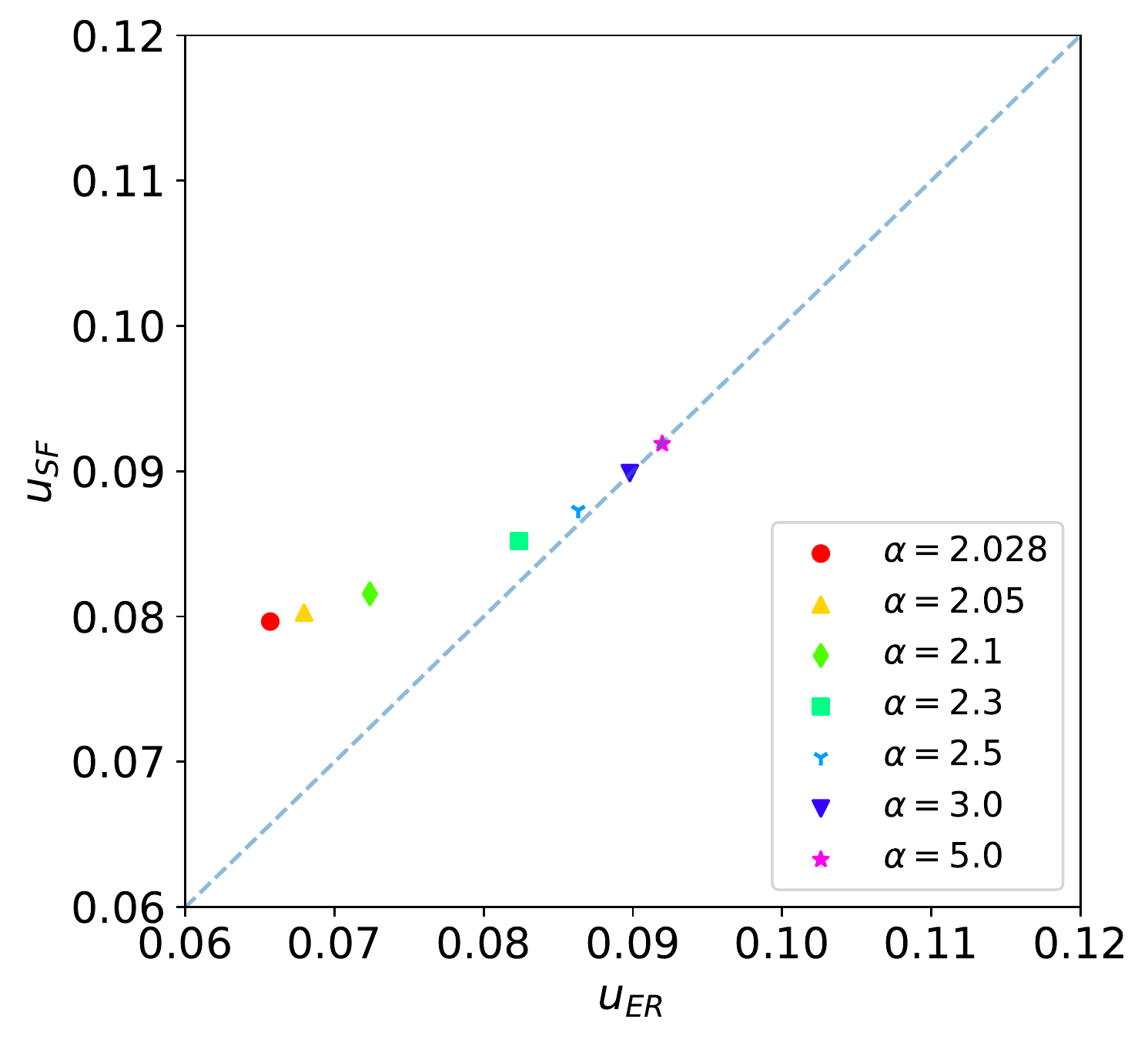}
            \subcaption{$u_{ER}$ v.s. $u_{SF}$}
            \label{fig:sf_vs_random_u}
        \end{minipage}
        \begin{minipage}{0.5\hsize}
            \centering
            \small
            \includegraphics[width = \hsize]{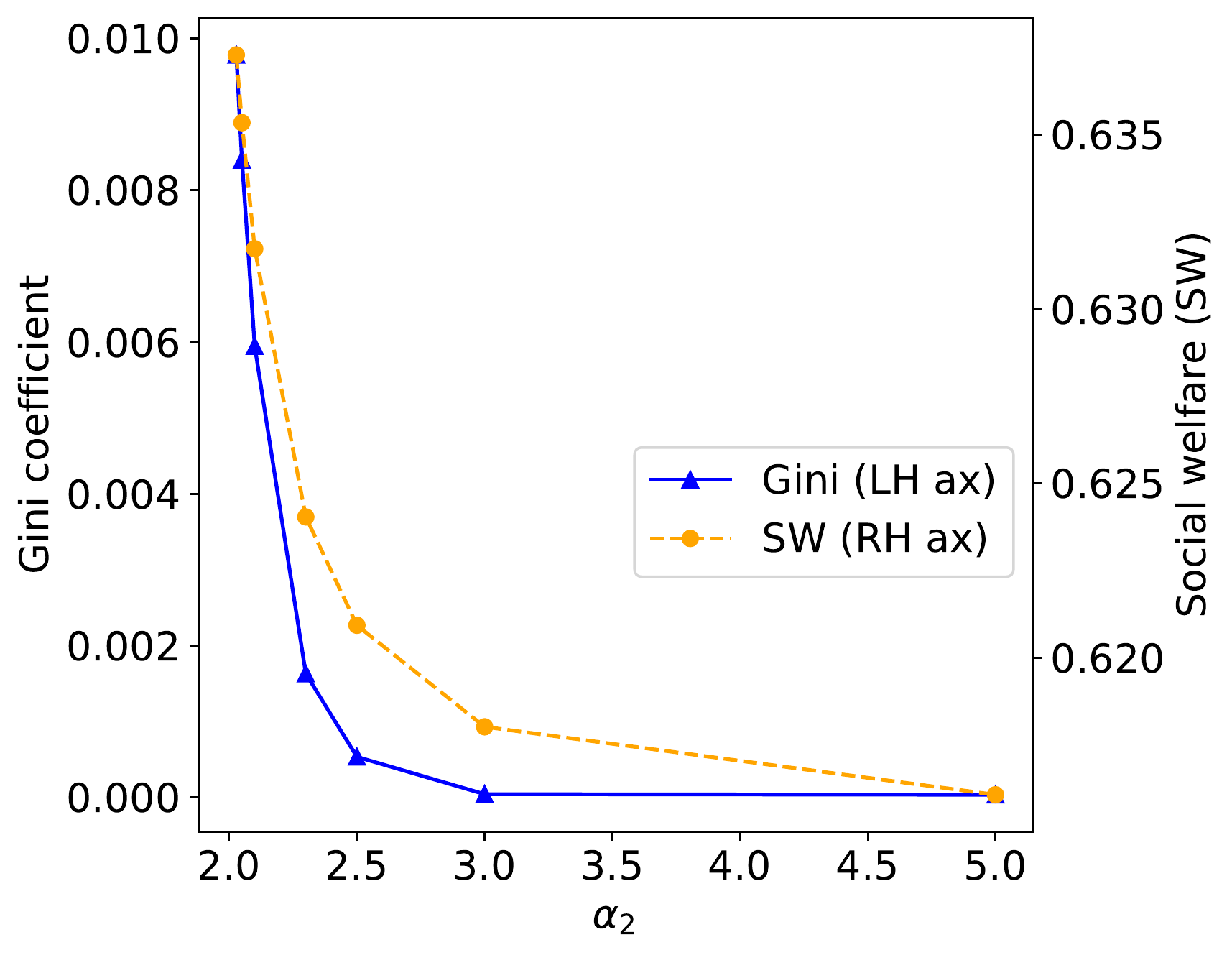}
            \subcaption{Gini coefficient and social welfare}
            \label{fig:sf_vs_random_G}
        \end{minipage}
        \caption{Comparison of unemployment rate, Gini coefficient, and social welfare. \ER network group ($ER$) v.s. scale-free network group ($SF$).}
        \label{fig:sf_vs_random}
        
        \centering
        \tabcaption{Expected degree for each $\alpha_2$}
            {\scriptsize
                \begin{tabular}{cccccccccccc} \hline
                     $\alpha_2$ & $2.028$ & $2.05$ & $2.1$ & $2.3$ & $2.5$ & $3.0$ & $5.0$ \\ \hline
                     $E_i [d_i]$ & $22.47$ & $12.86$ & $6.78$ & $2.74$ & $1.95$ & $1.37$ & $1.04$ \\ \hline
                \end{tabular}}
            \label{tab:exp_deg_alpha}
    \end{figure}

        We implement the following comparison between an \ER to a scale-free network group with an identical expected degree.
        Let group 1 and group 2 comprise an \ER and a scale-free network with scale parameter $\alpha_2$.
        The expected degree is calculated from the generated distribution of group 2.
        An \ER network is generated with the expected degree.
        Finally, we compute the endogenous variables at the steady state.
        Results are shown in Figure \ref{fig:sf_vs_random} and the expected degree associated with each $\alpha_2$ in Table \ref{tab:exp_deg_alpha}.
        If a point is on the 45 degree line in Figure \ref{fig:sf_vs_random_u}, two groups hold the equivalent unemployment rate.

        The scale-free network group gets a higher unemployment rate than the \ER network group under the small scale parameters.
        One intuition is rather simple.
        In a scale-free network with a low scale parameter, a few workers have many contacts while others have few contacts.
        Thus, workers in a scale-free network are more congested in information sharing than in an \ER network if the average number of friends is the same.
        The matching inefficiency arises in information sharing when too much information is gathered by a few popular workers.
        In most cases, the scale-free nature of social networks is worse for worker groups.
        

\section{Comparative Analyses} \label{sec:compara}

\subsection{Job Network Connections}

In this section, we carry out comparative analyses.
At first, we check the effect of job connectivity on inter-group inequality.
The previous literature usually considers the simple case with constant $d_f$.
If the number of connections in the job network increases, the probability that a referrer finds a vacant job position also increases.
We compute the equilibrium in different $d_f$ with two worker groups having the same expected number of contacts but different network structures, \ER and scale-free networks. 
We put $E_1[d_1] = E_2[d_2]=22.47$ and parameters in Table \ref{tab:calibration_parameter_fix} except for $d_f$.
Let the set of $d_f$ be $\{ 0,1,2,3,5,10,16,20,40 \}$.

\begin{figure}
        \begin{minipage}{0.43\hsize}
        \centering
        \small
        \includegraphics[width = \hsize]{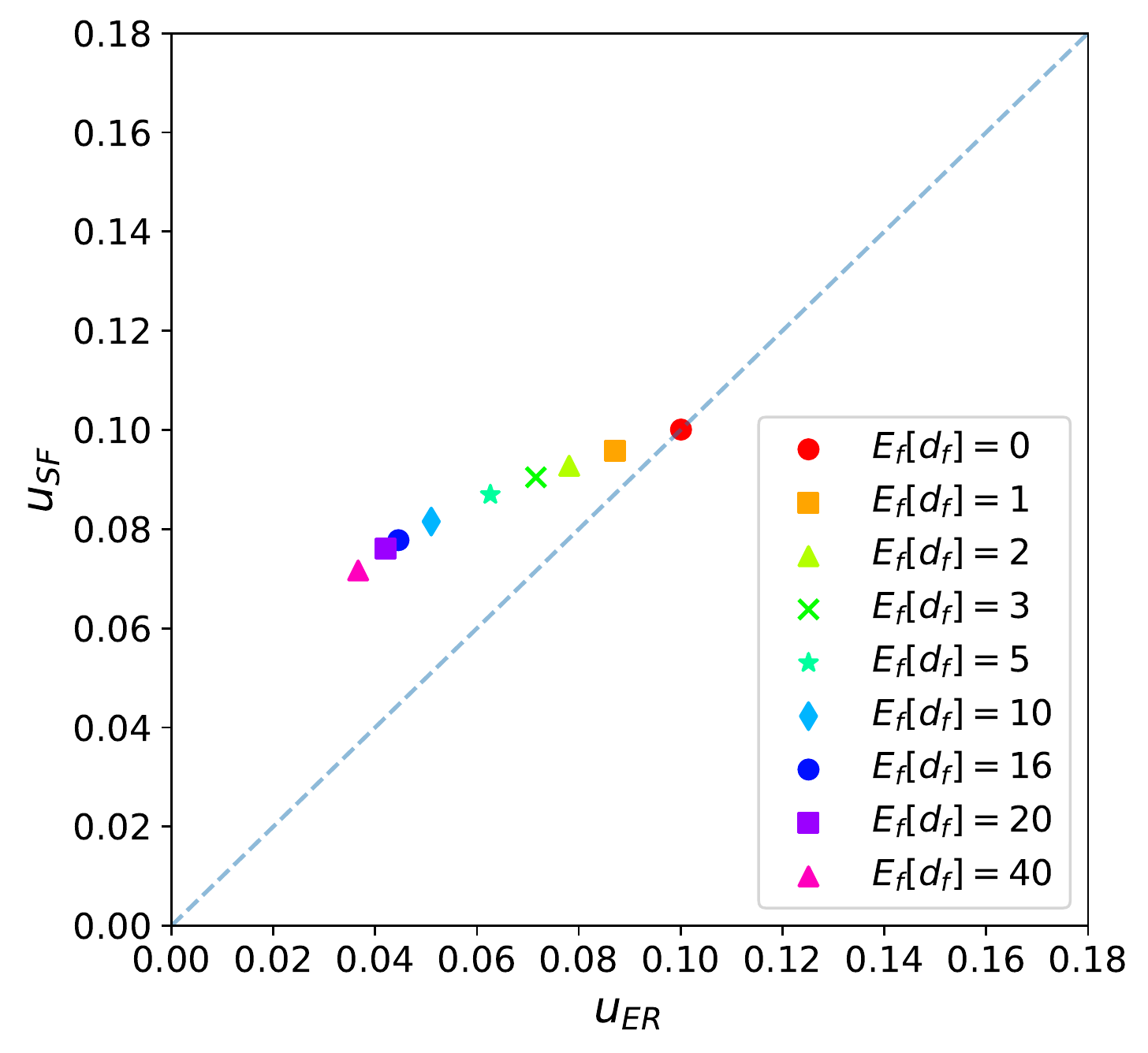}
        \subcaption{$u_{ER}$ v.s. $u_{SF}$}
        \label{fig:df_change}
        \end{minipage}
        \begin{minipage}{0.47\hsize}
        \centering
        \small
        \includegraphics[width = \hsize]{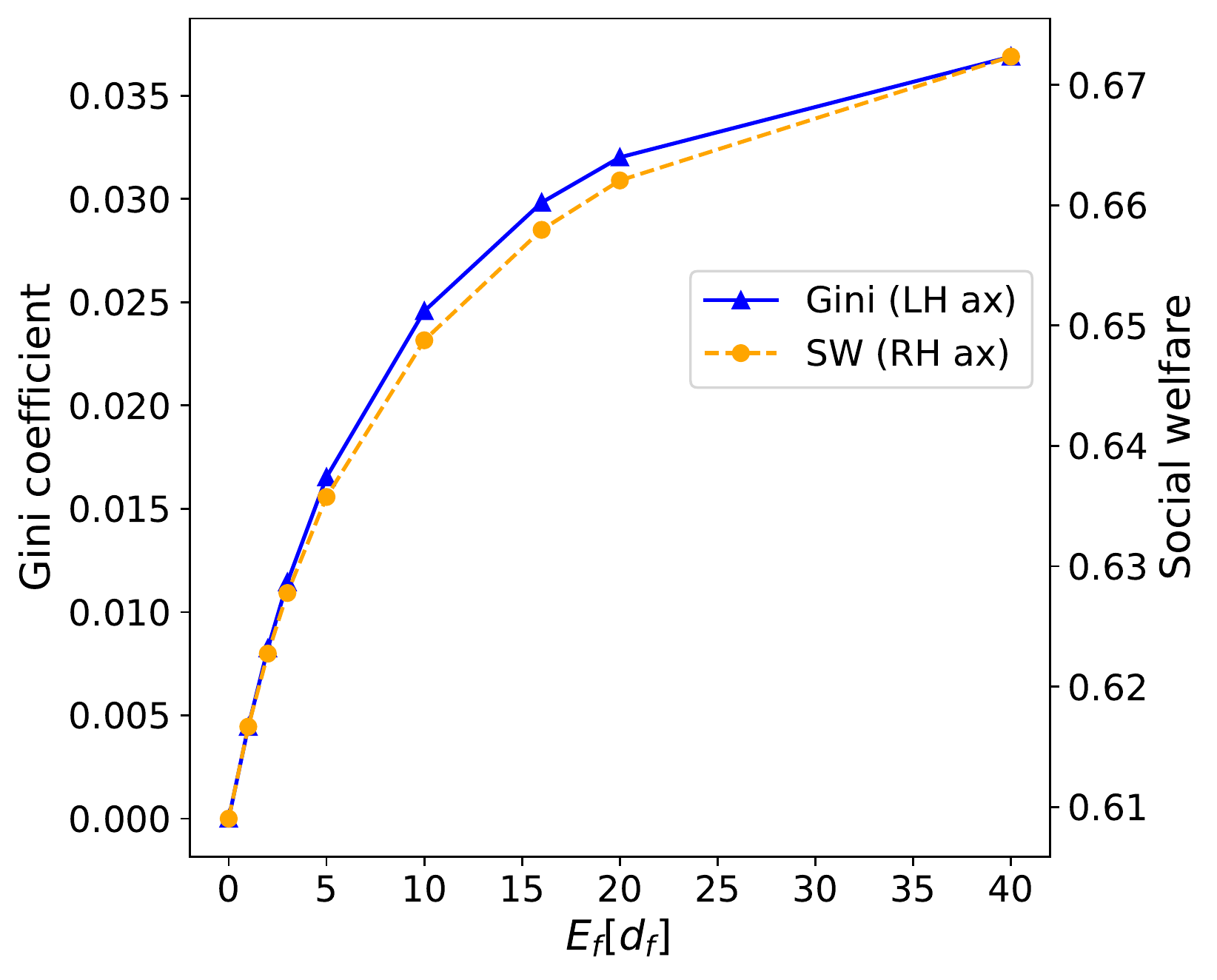}
        \subcaption{Gini coefficient and social welfare}
        \label{fig:df_change_GW}
        \end{minipage}
        \caption{Change in the number of connections in the job network, $d_f$. Two groups have the same expected degree $E_1[d_1] = E_2 [d_2] = 22.47$, while they have different network structures (\ER v.s. scale-free). }
\end{figure}

Results from calculations are shown in Figure \ref{fig:df_change} and \ref{fig:df_change_GW}.
As $d_f$ is increased from $d_f=0$ with complete equality, the unemployment rate worsens for the worker group with a scale-free social network structure, and the inequality increases monotonically in terms of the Gini coefficient.
Despite the expansion of inequality, social welfare improves along with the $d_f$ since referral hiring becomes available for many workers, leading to a large match surplus.
However, most additional benefits from referrals go to one group with a better social network structure for job matching.

\subsection{Policy Intervention to Referral} 
    
    Next, we examine the effects of referral control by the policymaker.
    Suppose that the policymaker can control $\phi$ by regulations or announcements.
    We employ the parameter set of Table \ref{tab:calibration_parameter_fix} except for $\phi$.
    At first, let group 1 and group 2 form an \ER and a scale-free network, respectively.
    We set expected degree such that $E_1[d_1] = E_2 [d_2] = 22.47$ following before discussion.
    Scale parameter of group 2 is calibrated to $\alpha = 2.028$.
    Calculations are carried with $\phi \in \{0, 0.001, 0.01,0.1, 0.408, 0.1, 0.3, 1.0\}$.
        
    \begin{figure}
        \begin{minipage}{0.43\hsize}
        \centering
        \small
        \includegraphics[width = \hsize]{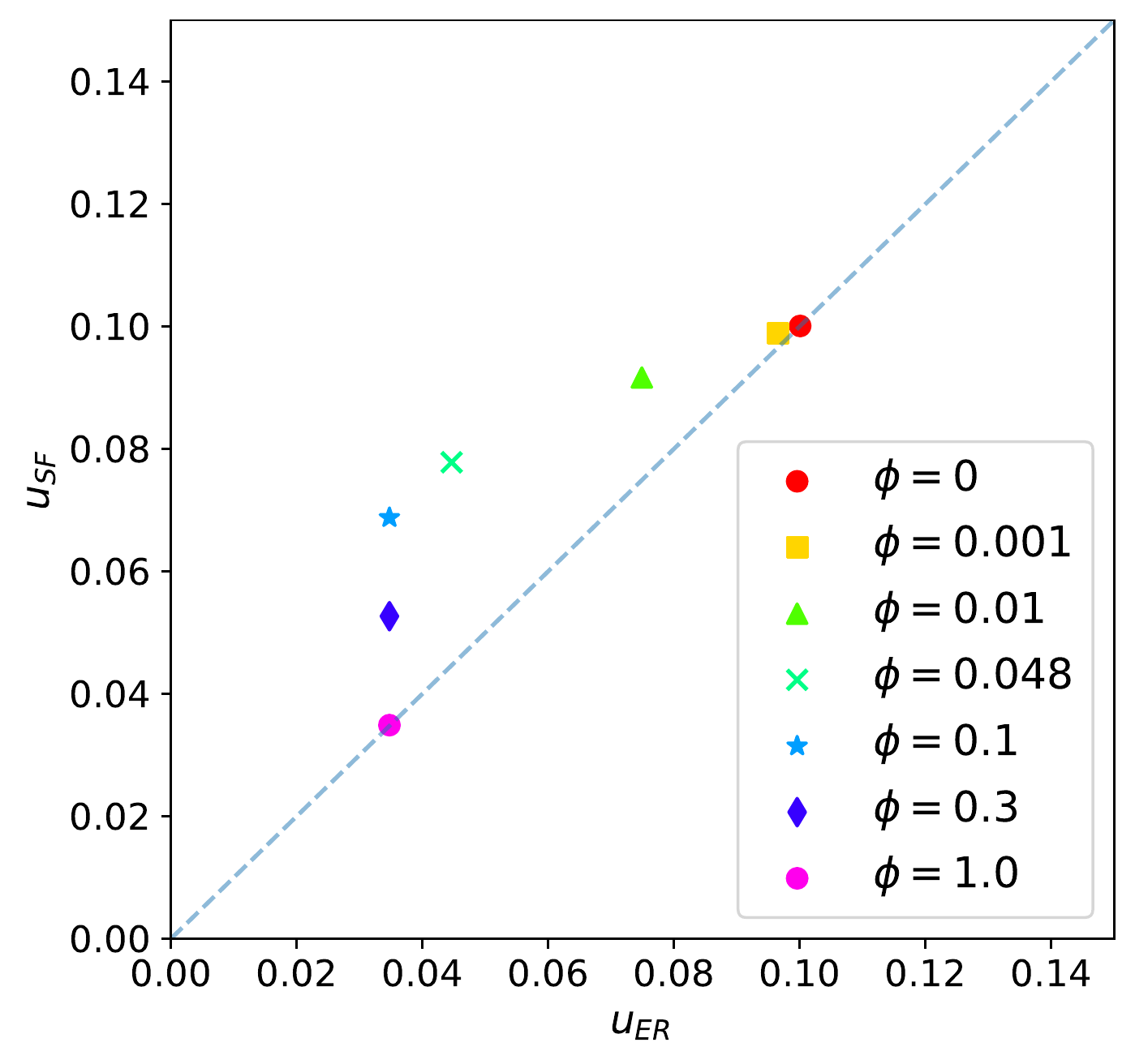}
        \subcaption{$u_{ER}$ v.s. $u_{SF}$}
        \label{fig:phi_random}
        \end{minipage}
        \begin{minipage}{0.47\hsize}
        \centering
        \small
        \includegraphics[width = \hsize]{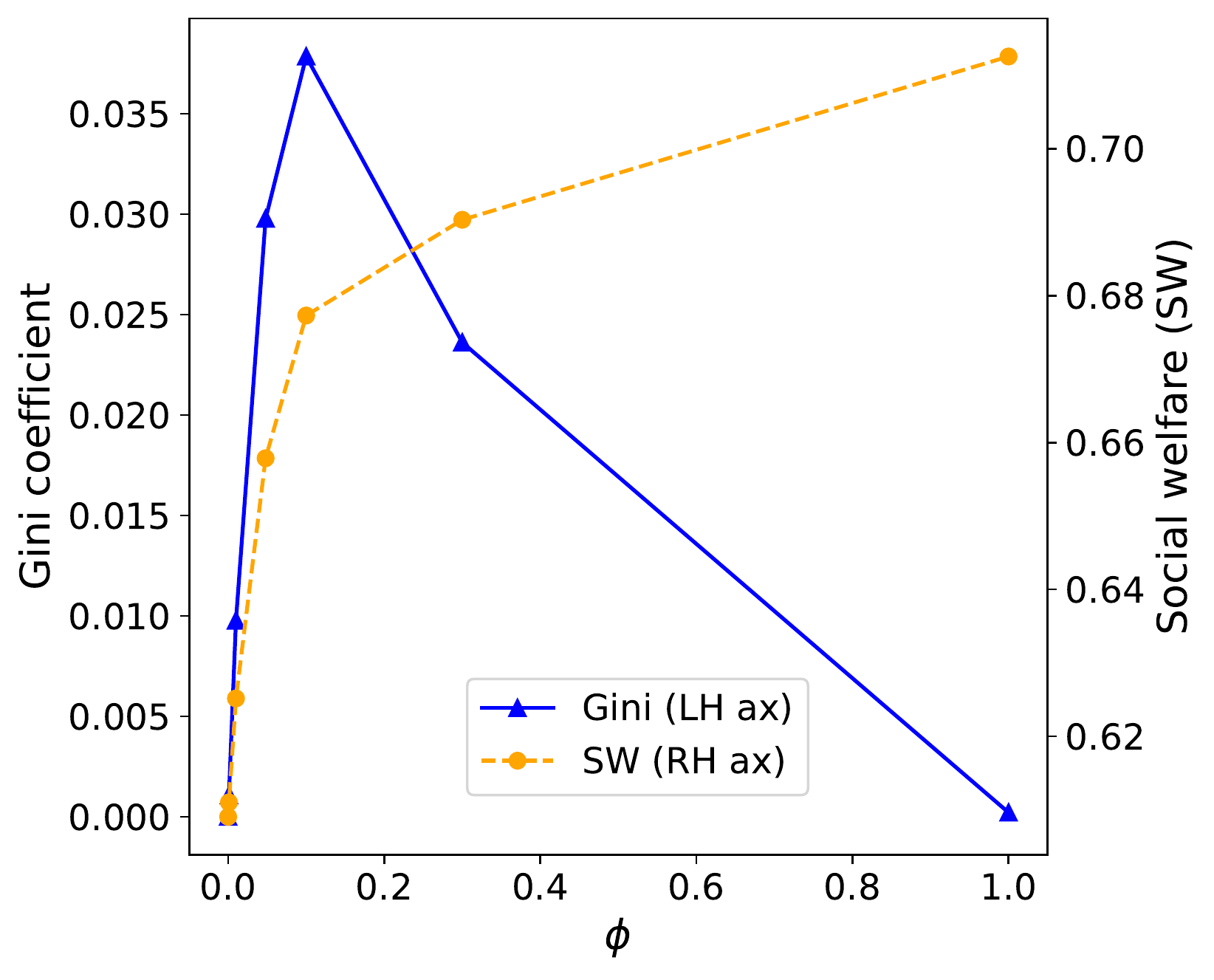}
        \subcaption{Gini coefficient and social welfare}
        \label{fig:phi_random_GW}
        \end{minipage}
        \caption{Policy intervention to referral prevalence, $\phi \in [0,1]$. Two groups have the same expected degree $E_1[d_1] = E_2 [d_2] = 22.47$, while they have different network structures (\ER v.s. scale-free). }
    \end{figure}
        
        We show the results in Figure \ref{fig:phi_random} and \ref{fig:phi_random_GW}.
        When $\phi = 0$, both groups cannot use referral, regardless of the worker group.
        Then, the model is identical to the basic Pissarides type model.
        Naturally, there is no inequality between groups.
        Moving focus onto the Gini coefficient and social welfare in Figure \ref{fig:phi_random_GW}.
        Although the welfare increases monotonically in $\phi$, the Gini coefficient does not.
        While the inequality increases at first, the gap between the two takes a maximum value around $\phi = 0.1$ and decreases from that point.
        If the economy is in the regime such that $\phi >0.1$, restrictions of referral not only make the welfare worse, but also aggravate the inequality. 
        The result is consistent with the discussion in \cite{igarashi2016distributional}.

\section{Conclusion} \label{sec:conclusion}
    
    In this paper, we investigated whether network structure can affect the inequality of unemployment and wage rate between groups.
    Hence, we constructed a general equilibrium model with referral hiring and analyzed the effect of the characteristics of network structure on inequality.
    We have found that not only the average number of contacts in the group, but also the social network structure has a significant impact on the inequality between groups.
    Random regular networks have an advantage in lowering unemployment rates and increasing wage rates over \ER networks, although the gap is not large.
    However, scale-free networks could be disadvantageous for worker groups and the effects cannot be negligible.
    Scale-free network structures in social networks are more congested than other network structures.
    This result suggests that network structures need to be considered when connecting theoretical models and empirical data because empirical findings show that a lot of social networks are scale-free.

    Also, we have shown comparative analyses including expansion of job network connectivity and effects of policy intervention via the referral prevalence.
    In the first case that the number of connections in the job network increases, social welfare improves while the inequality deteriorates monotonically.
    In the second case, there is a non-monotonic relationship between the frequency of referral and inequality.
    If the government restrains referral hiring when it is sufficiently widespread, inequality between worker groups may widen.
    
    The analysis could be extended in several directions for the next step.
    First, we could use actual data to identify the extent of the gaps between different social groups attributed to differences in the network structures.
    The model presented in the paper needs scarce information on the network for the empirical analysis because it only needs the degree distribution's information in worker groups.
    Second, we could incorporate time-dependent networks to search and matching model with overlapping generations \citep[e.g.][]{cheron2011age,cheron2013life,fujimoto2013note,hahn2009search}.
    Social networks usually change over time and are called \textit{temporal networks} \citep{holme2019temporal}.
    If the change of the networks along with generations can be captured, we can build temporal networks in the model to analyze intergenerational inequality.
    Finally, qualitative differences in networks, such as clustering coefficients, could be incorporated into the model. 
    In the model, the probability of a clustering coefficient being positive is almost zero since we use configuration models in fact \citep[][]{catanzaro2005generation,newman2018networks}.
    Also, there are other qualitative variables to characterize a network property, for example, average distance and diameter \citep{chung2002average}.
    By incorporating the differences in these qualitative variables in the analysis, we could examine the extent of changes that affects the economy.
    


\bibliographystyle{econ-aea}

\end{document}